\documentclass[%
reprint,
amsmath,amssymb,
aps,
]{revtex4-2}

\usepackage[caption=false]{subfig}
\usepackage{braket}
\usepackage{graphicx}
\usepackage{dcolumn}
\usepackage{bm}
\usepackage{xcolor}

\begin{document}
\preprint{APS/123-QED}
	
\title{Critical temperature of the classical $XY$ model via autoencoder latent space sampling}	
\author{Brandon Willnecker}
\author{Mervlyn Moodley}
\affiliation{School of Chemistry and Physics, University of KwaZulu-Natal, Westville Campus, Private Bag X54001, Durban, 4000, South Africa}
	
\date{\today}
	
\begin{abstract}
The classical $X$Y model has been consistently studied since it was introduced more than six decades ago. Of particular interest has been the two-dimensional spin model's exhibition of the Berezinskii–Kosterlitz–Thouless (BKT) transition. This topological phenomenon describes the transition from bound vortex-antivortex pairs at low temperatures to unpaired or isolated vortices and anti-vortices above some critical temperature. In this work we propose a novel machine learning based method to determine the emergence of this phase transition. An autoencoder was used to map states of the XY model into a lower dimensional latent space. Samples were taken from this latent space to determine the thermal average of the vortex density which was then used to determine the critical temperature of the phase transition.
\end{abstract}
	
\maketitle
\section{\label{sec:s0}Introduction}	
\noindent Besides its application to a plethora of fields in the physical sciences \cite{CCC}, machine leaning based techniques has had a profound influence in the study of condensed matter systems \cite{MLCM}. Of importance in these systems is the characterisation of phases of matter. Recently, Carrasquilla and Melko \cite{CM} used a simple supervised learning approach to identify phase transitions and almost simultaneously, Wang \cite{Wang} proposed unsupervised learning techniques for discovering phase transitions in many-body systems. In the latter's work, the order parameter and structure factor was used as indicators of phase transitions. Since then, there has appeared numerous papers on using machine learning to identify and classify phase transitions, including topological phase transitions\cite{Loop,Khan,Nieva,Beach,Rem, Zhang} which proves to be more difficult since these are defined in terms of non-local properties.

There are several existing machine learning methods for studying the Berezinskii–Kosterlitz–Thouless (BKT) transition in the $XY$ model. Zhang, Liu, and Wei \cite{Zhang} used supervised machine learning to determine the phase boundary. A fully connected neural network was trained on Markov chain Monte Carlo (MCMC) samples generated at temperatures before, near and after an estimated critical temperature, $T_c$. Once trained, this model was able to identify the transition temperature based on the switching in the models predictions. Ng and Yang \cite{Ng} also used autoencoders in their study of the classical XY model as we did in this paper. However, they used the mean-square-error loss function as a measure of the disorder in the given system. Phase transition points (including first-order, second-order and topological ones) could be detected by the peaks in the standard deviation of the loss function. Shiina {\textit{et al.}}\cite{Shiina} adopted a similar technique to \cite{Zhang} but instead of using the spin configurations, they utilized long-range correlations, $g_i(r) = s_i s_{i+r}$, as the inputs to a fully connected neural network. Again, the switching in the predicted output was used to determine the transition temperature. Miyajima and Mochizuki \cite{yusuke} proposed two machine learning methods for the detection of phase transitions in Heisenberg, Ising and $XY$-like models. They first used a supervised learning technique similar to \cite{Zhang} whereby inputs are labeled according to there phase. Once the neural network is trained, it can be sampled near the $T_c$ point. This point can be determined once the neural network's output changes (ie. a phase transition has been detected). The second method is a temperature prediction neural network. The input is a spin configuration $\vec{s}=(s_1..s_{L\times L})$ for an $L\times L$ lattice and the output is a 200 dimensional dimensional vector, $\vec{o} = (o_1,..o_{200})$, where each entry $o_n$ is a probability that the spin configuration is at temperature $T_n=n\Delta T=n0.01J$ where $J$ is the coupling strength of the spin-spin interaction. The phase changes are detected by studying the distinct patterns in heat maps of the weights of the neural networks for $T<T_c$ and $T>T_c$. The change in the pattern indicates a change in the phase.

Instead of training a neural network to predict the phase of a given state, we propose a method to more efficiently sample the space of states. We train an autoencoder to map the large space of states to a lower dimensional latent space. This latent space may be much smaller but each element still contains the required information to reconstruct the original state. We can then sample from this space in constant time to calculate certain quantities that show a phase transition has occurred.

The paper is organized as follows. In Sec.~\ref{sec:s2} we provide the details for how the autoencoder works and how the latent space samples are taken. In Sec.~\ref{sec:s3} we revise the classical XY lattice model and explain why it is advantageous to study the continuous analogue, $\theta(x,y)$, with local $U(1)$ symmetry removed. This is done by introducing an auxiliary field $A(x,y)$ that is derived from the continuous $\theta(x,y)$ field. This field is analogous to the average energy of a spin and its neighbours.  In Sec.~\ref{sec:s4} we use the concept of vortex density to determine at which temperature, $T_c$, after which the vortices in the XY model become unbounded. The conclusion follows in Sec.~\ref{sec:s5}.
\newpage
\section{\label{sec:s2}Autoencoders}
An autoencoder is a type of neural network that is used in supervised learning to provide efficient codes (compressed representation) to unlabelled input data \cite{auto_endoder_info, auto_encoder_use}. An autoencoder is made of two sections called the encoder and decoder with a ``bottle neck'' in between as can be seen in Fig. \ref{fig:encoderdecoderdigram} below.
\begin{figure}[h!]
	\centering
	\includegraphics[width=\linewidth]{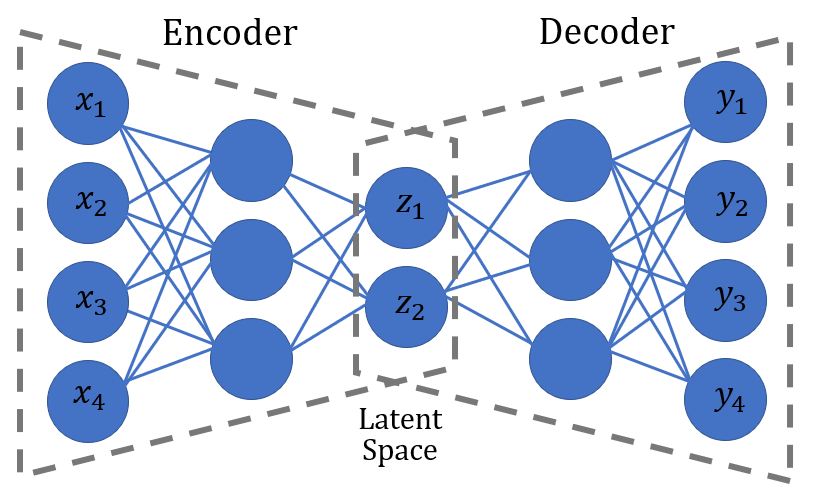}
	\caption[Schematic of an autoencoder]{The input vector $(x_1,x_2,x_3,x_4)$ is compressed into a lower dimensional latent space representation vector $(z_1,z_2)$ by the encoder. The decoder then attempts to reconstruct the original vector, ie. the vector norm $|x-y|$ is minimal for all input vectors $x$.}
	\label{fig:encoderdecoderdigram}
\end{figure}

\noindent This can be described mathematically by the function
\begin{equation}
	f:X\rightarrow Z
\end{equation}
which embeds the vectors from $X$ into a lower dimensional space $Z$ and the function
\begin{equation}
	g:Z\rightarrow X
\end{equation}
which reverses the action of $f$. The aim is for the autoencoder to ``learn'' these functions such that
\begin{equation}
	\forall x\in X, ~~(g\circ f)(x) = x
\end{equation}
\noindent The neural network is trained through back propagation in order to minimize a loss function. 
\noindent A typical loss function for an autoencoder is given by
\begin{equation}
	L=\sum_{i=1}^{n}\frac{1}{2}(y_i-x_i)^2
\end{equation}
where $x_i$ are the inputs and $y_i$ are the reconstructed output \cite{loss_fucntion, mnielson}. This loss functions ensures that the neural network correctly reconstructs the inputs by minimizing the difference between them. The narrowing of the network to a ``bottle neck'' is essential for the neural network to learn the required compression. The set of vectors produced from this compression, $Z$, is called the latent space. This latent space is a lower dimensional representation of the input space.\\\\
\newpage
\section{\label{sec:s3}The Model}
\begin{figure}[h!]
		\centering
		\includegraphics[width=\linewidth]{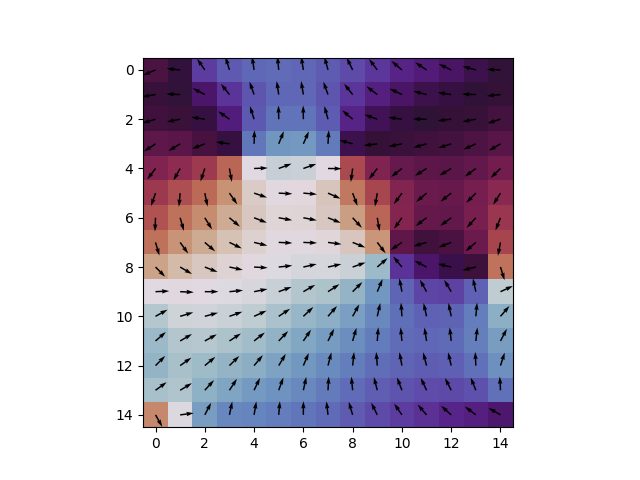}
		\caption[Illustration of vortex and anti-vortex]{Here we have examples of vortex (middle right) and anti-vortex (upper left) configurations on the lattice. In the continuum limit, these become the singular solutions (topological defects) mentioned above. The background colour shows the $\theta(x,y)$ field.}
		\label{fig:vortexantivortex}
	\end{figure}

The classical XY model is a lattice model in which each site is occupied by a two dimensional unit vector ${\vec s}=(\cos\theta,\sin\theta)$. The configuration, $S=\{\vec s_i\}$, is an assignment of each ${\vec s_i}$, or equivalently, an angle $\theta\in[-\pi,\pi]$ to each lattice site. The total energy of the configuration is given by the Hamiltonian
\begin{equation} \label{H1}
H=-\sum_{i\not=j}J_{ij}{\vec s_i}\cdot {\vec s_j} - \sum_{i}{\vec h_i}\cdot {\vec s_i}
\end{equation}
where $J_{ij}$ is the strength of interaction between the $i^{th}$ and $j^{th}$ site, and ${\vec h_i}$ is a site dependent external field. For our purposes, we will use a simplified version of the Hamiltonian, however it should be noted that the general case can be handled in a similar way. We will make three simplifying assumptions. Firstly, the strength of interaction will be taken as site independent. Secondly, we will not include an external field and lastly, we will only consider nearest neighbour interactions. Eq.~(\ref{H1}) will therefore read
\begin{equation}\label{H2}
	H = -J\sum_{<i,j>}\cos(\theta_i-\theta_j),
\end{equation}
where summation is over nearest neighbours. The angles $\theta_i$ and $\theta_j$ are the angles of the vectors $s_i$ and $s_j$ respectively. The Mermin–Wagner theorem \cite{Hohenberg_Mermin_Wagner_theorem} states that continuous symmetries cannot be spontaneously broken at finite temperature. The fact that this theorem does not apply to discrete symmetries was seen previously in the 2D Ising model. Since the XY model has a continuous symmetry ($\theta_i\rightarrow \theta_i+\delta\theta$), we do not expect a typical phase transition. Instead, we see a topological phase transition known as the Berezinskii–Kosterlitz–Thouless transition \cite{xy1, xy2, xy3, xy4}. This transition can be studied by first taking the continuum limit of the lattice model. The continuum Hamiltonian is given by
\begin{equation}\label{En}
	H(\theta)=\int\frac{J}{2}(\nabla\theta)^2dxdy,
\end{equation}
where the field $\theta$ replaces the discrete angle assignments $\theta_i$. The field configurations that give stationary $H$ can be found using
\begin{equation}
	\frac{\delta H}{\delta\theta}=0\implies \nabla^2\theta=0,
\end{equation}
which give two solutions. The first solution is the uninteresting ground state given by $\theta(x,y)=\text{constant}$ and the second, more interesting, solution involves the addition of vortices and anti-vortices which are topological defects in the $\theta$ field. These vortices are singular solutions to the equation
\begin{equation}
	\nabla^2\theta = 0
\end{equation}
with
\begin{equation}\label{li}
	\oint_C \nabla\theta\cdot dl=2\pi q, q\in\mathbb{Z}
\end{equation}
\noindent The integral is taken around a closed loop surrounding the singular point of the vortex. This integral gives an integer multiple of $2\pi$ because the net change in the spin vector must be some multiple of a full revolution. The integer $q$ is the ``charge'' of the vortex/anti-vortex. Vortices have charge $+1$ and anti-vortices have a charge of $-1$. Illustrations of these voticies and anti-vortices as they would appear on a lattice are shown in Fig. \ref{fig:vortexantivortex}.

\noindent To calculate the energy of a vortex, we first note that the angular symmetry of $\theta$ allows us to write $\theta=\theta(r)$. We can then use Eq.~({\ref{li}) to find $|\nabla\theta|$.
\begin{equation}
	2\pi q = \oint_C\nabla\theta(r)\cdot dl = |\nabla\theta|2\pi r \implies |\nabla\theta|=\frac{q}{r}
\end{equation}
Eq.~(\ref{En}) can be used to calculate the energy of a vortex/anti-vortex. This gives
\begin{equation}
	E = \frac{J}{2}\int(\nabla\theta)^2dxdy = Jq^2\pi\ln\Big(\frac{L}{a}\Big),
\end{equation}
where $L$ is length of the system and $a$ is a lower cut-off value that can be taken as the lattice spacing from the original problem. This energy diverges in the thermodynamic limit so we do not have single vortex or single anti-vortex excitations. Instead, dipoles consisting of vortex and anti-vortex pairs can exist since they have finite energy. This is due to the fact that a closed loop surrounding the dipole contains no charge, $q_\text{net}=(+q)+(-q)=0$.

As already stated, there is no spontaneous symmetry breaking at finite temperature, however, there is a transition between long-range correlations at low temperature and short range correlations at high temperature. Kosterlitz and Thouless\cite{xy3} showed that at low temperatures the vortices occur in tightly bound pairs. As the temperature increases past a transition point, $k_BT_{KT}/J\approx0.893$, the pairs undergo deconfinement which results in a change in the order parameter from a power-law to exponential.
	
\section{\label{sec:s4}Generating and sampling the latent space}
We investigate the unbinding phenomena by studying the density of these vortices (number of vortices per unit area) as a function of temperature. We expect that the vortex density is almost zero for low temperatures and then increases after the transition temperature. In order to calculate the thermodynamic average of the vortex density, we can generate samples from the associated Boltzmann distribution \cite{vorext_density_sampling}. This can be rather computationally expensive. We instead use an autoencoder, illustrated in Fig. \ref{fig:xyautoencoder}, to generate a lower dimensional latent space from the full configuration space. We can then easily sample points from this latent space, pass these points through the decoder and thus generate as many field configurations as we need.
\begin{figure}[ht!]
	\centering
	\includegraphics[width=\linewidth]{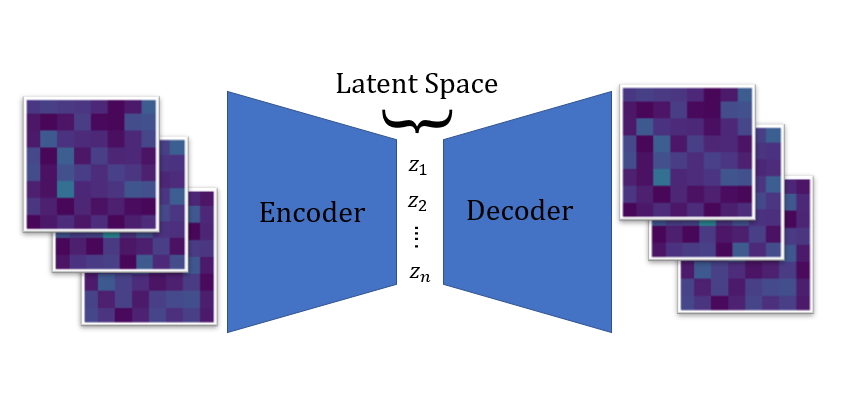}
	\caption[XY model auto-encoder]{The autoencoder was designed with a 60-dimensional latent space. The architecture and training details are elaborated on in the appendix.}
	\label{fig:xyautoencoder}
\end{figure}

 However, more work needs to be done if we simply use $\theta(x,y)$ field configurations. This is because these fields have internal $U(1)$ symmetry which needs to be taken into account. We can either design a neural network that would respect this symmetry or we could over specify the training data in order for the neural network to learn the symmetry from examples \cite{nn_symmetry, nn_symmetry_2}. Instead of doing either of these, we propose the introduction of an auxiliary field, $A(x,y)$, that removes the symmetry from the $\theta(x,y)$ fields. We define the auxiliary field $A(x,y)$, given $\theta(x,y)$, by
\begin{equation}
	A(x,y)=\frac{1}{\sigma \sqrt{8\pi}}\int_{D} N(u,v)\Big[1-\cos\Big(\theta(x,y)-\theta(u,v)\Big)\Big]dudv
\end{equation}
where $D(x,y)$ is a disk of radius $\sigma$ centred at $(x,y)$ and $\sigma$ is a length on the order of the width of a vortex/anti-vortex. This field quantifies the average variation of $\theta(x,y)$ around a local neighbourhood of radius $\sigma$. \newpage\noindent $N(u,v)$ is a Gaussian centred at $(x,y)$. This term ensures that only field values close to $(x,y)$ are considered in the averaging process. The cosine term is analogous to the term in Eq.~(\ref{H2}) and is used to remove the unwanted symmetry. This term also ensures that the field is bounded which will be important when implementing the autoencoder. An example of this $\theta(x,y$) field and its corresponding $A(x,y)$ field are illustrated in Fig. \ref{fig:A_field}. Vortices and anti-vortices will have large field variations in their neighbourhood and so will result in a large field value. Regions with no vortices or anti-vortices will have little to no variation in the field. This will result in a very small field value. We can thus characterize the vortex/anti-vortex density by the magnitude of $A(x,y)$ across the extent of the field. The autoencoder is then trained using these fields that are derived from $\theta(x,y)$ fields generated using a standard MCMC method which is explained in the appendix. Other configuration generation methods such as those by Swendsen and Wolff \cite{method_swendsen, method_wolff} can be used to generate the training data. However, as long as a sufficient amount of training samples is generated, the precious method of generation is not important.
\begin{figure}[ht!]
	\centering
	\includegraphics[width=\linewidth]{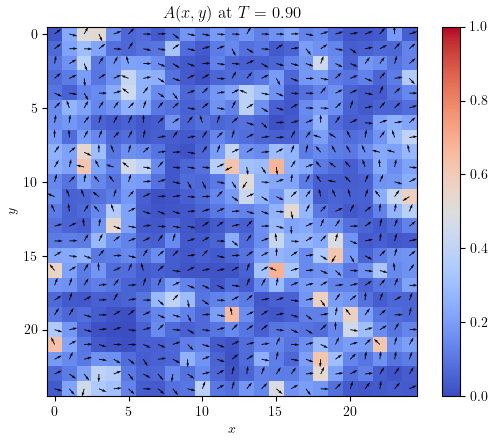}
	\caption{This $\theta(x,y)$ field was generated using MCMC at $T=0.9$. The background is the auxiliary $A(x,y)$ field calculated using Eq. 13.}
	\label{fig:A_field}
	\end{figure}

\noindent Once trained, the latent space was analysed by passing in $A(x,y)$ fields and generating histograms from the produced latent space values for each dimension of the latent space. These histograms were then used to determine the mean and standard deviation of the distribution for each latent space dimension.
\noindent New fields can be generated by passing in a sampled latent space vectors, $z=(z_1,z_2,z_3,...,z_{60})$, into the decoder portion of the autoencoder. Each $z_i$ is sampled from the respective Gaussian with mean $\mu_i$ and standard deviation $\sigma_i$. The particular Gaussian distribution $N(\mu_1=0.14,\sigma_1=0.17)$ for the latent space dimension $z_1$ is shown in Fig. \ref{fig:z1gaussian} below. 
\begin{figure}[ht!]
	\includegraphics[width=\linewidth]{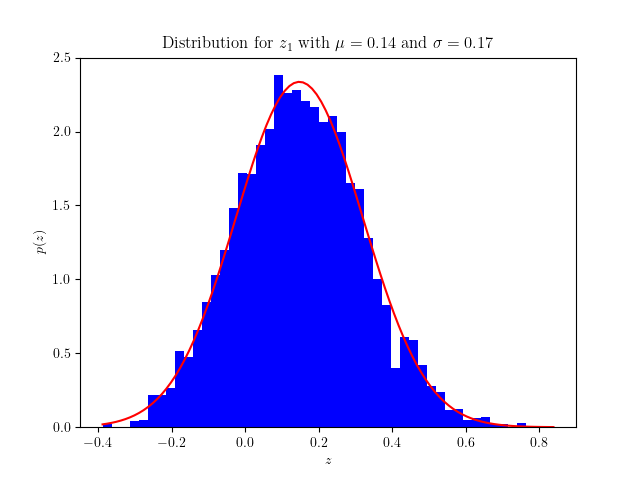}
	\caption{Histogram of $z_1$ values sampled from the first latent space dimension. Similar distributions can be obtained for the other dimensions of the latent space as explained above.}
	\label{fig:z1gaussian}
\end{figure}

This method is similar to, but technically different from, a variational autoencoder \cite{intro_to_vae}. Through the training process, an autoencoder learns to map a high dimensional vector to a lower dimensional latent space. The exact process is not known explicitly but is manifested in the learned parameters. On the other hand, a variational autoencoder first maps the input to a mean vector and a standard deviation vector of a predefined distribution. This distribution is then sampled to produce the latent space vector. The generation of these latent space distributions can be done using either method however it should be noted that you do not need to predefine a distribution over the latent space before training when using an autoencoder. The latent space sampling process is very computationally inexpensive compared to the standard MCMC algorithm. These sampled fields can then be used to calculate the thermodynamic average of the vortex density at a given temperature. We cannot directly count the number of vortices so instead we use the average,
\begin{equation}
	\braket{A} = \frac{1}{L^2}\int_{0}^{L}\int_{0}^{L}A(x,y)dxdy,
\end{equation}
as a proxy. $A(x,y)$ is analogous to the energy at and around the point $(x,y)$. If each ``vortex'' has energy $\epsilon$ then $\frac{1}{\epsilon}\braket{A} = n$ gives the average number of unbound votices over the extent of the field. Ideally, we would expect the function $\braket{A}$ to be
\begin{equation}\label{A(T)}
	\braket{A} =\Big\{
	\begin{array}{lr}
		0 & \text{if } T < T_c\\
		a(T-T_c) & \text{if } T \ge T_c
	\end{array}
\end{equation}
where we have vortex unbinding above the critical temperature. The number of vortices, and hence vortex density, will then grow linearly with temperature. In practice, finite size effects and finite sampling effects will result in an approximate form of $\braket{A}$ as shown in the figure below.
\begin{figure}[h!]
	\includegraphics[width=\linewidth]{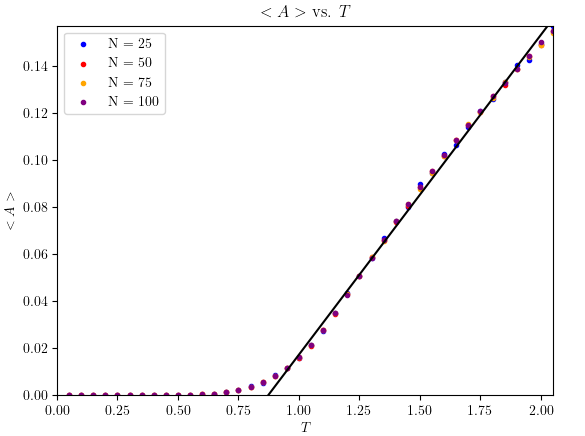}
	\caption{We calculate $\braket{A}$ as a function of $T$ over the temperature range $0 < T < 2$. This was done for field sizes $N=25, 50, 75$ and 100. The straight line in black is a line of best fit.}
	\label{fig:A_vs_T}
\end{figure}

\noindent Note the approximate fitting to Eq.~\ref{A(T)}. We can use the intercept point of the line of best fit as an approximation for $T_c$. The results are summarized in the table below.
\begin{table}[ht!]
	\begin{tabular}{|l|l|l|}
		\hline
		$N$ & $T_c$ Estimate & Error $\Delta T$ \\ \hline
		25  & 0.872833       & 0.020067         \\
		50  & 0.872213       & 0.020687         \\
		75  & 0.871779       & 0.021121         \\
		100 & 0.872756       & 0.02144          \\ \hline
	\end{tabular}
	\caption{Table of critical temperature estimates and associated errors for varying system sizes.}
\end{table}
\section{\label{sec:s5}Conclusion}
It was shown that an autoencoder can be a useful tool in reducing a given configuration space to a lower dimensional latent space that may be much easier to sample from. In this case, it was found that the latent space could be sampled using various Gaussian distributions. This latent space was sampled to calculate the thermal average of the vortex density and from this, one could determine the critical temperature ($T_{KT}$) at which this vortex density becomes non-zero. This method of latent space sampling is very general and so can be applied to many systems. The only requirement is that the systems need to contain a large enough amount of correlation between its constituents in order for the autoencoder to learn how to compress it with little loss. A large amount of correlation results in a large amount of redundant information can be removed during the compression stage of the autoencoder. With this in mind, one can extend this method to systems of large particles with solid-liquid phase transitions or systems with topological phase transitions \cite{top_phase_trans} since one does not need any order parameters during the training process. Systems with very little to no correlations (like an ideal gas) cannot be mapped to a lower dimensional space since the knowledge of the behaviour of part of the system gives no information on the behaviour of another part, ie. the specification of the system cannot be reduced. Future work for this method includes extensions to other learnable features such as magnetization where a magnetization density vector field $M(x,y)$ is derived from the $\theta(x,y)$ field instead of the auxiliary $A(x,y)$ field as done in this paper. Other work includes extensions to the generalized XY model which includes fractional vorices. In that case, a new vortex counting method needs to be implemented.
\section{\label{sec:6}Appendix}
\subsection*{Pseudo code for MCMC generation}
\begin{verbatim}
	#Generate a field of size N x N at temp. T
	TWO_PI = 6.283185
	N = 100
	T = 0.6
	STEPS = 1000 #Number of MCMC steps
	field[N][N]  #Field configuration
	
	#Initialize the field randomly
	#rand() returns a uniform random float on (0,1)
	for i = 0 to N
		for j = 0 to N
			field[i][j] = rand() * TWO_PI
	
	for step = 0 to STEPS
		#Choose a random site
		i = rand() * N; j = rand() * N
		E_prev = -cos(field[i][j]-field[i+1][j])
		         -cos(field[i][j]-field[i-1][j])
		         -cos(field[i][j]-field[i][j+1])
		         -cos(field[i][j]-field[i][j-1])
		#Propose a small change to the field
		delta = rand() * TWO_PI * 0.1
		field[i][j] += delta
		E_new = -cos(field[i][j]-field[i+1][j])
		        -cos(field[i][j]-field[i-1][j])
		        -cos(field[i][j]-field[i][j+1])
		        -cos(field[i][j]-field[i][j-1])
		dE = E_new - E_prev
		#Accept or Reject using Metropolis-Hastings
		p = min(1.0, exp(-dE/T))
		if rand() < p #accept
		else #reject
			field[i][j] -= delta
\end{verbatim}
\noindent The $\theta(x,y)$ and $A(x,y)$ fields were discretized into lattices of sizes N = 25, 50, 75 and 100. The size of the autoencoder input layer for each lattice is $N^2$. The size of each subsequent layer is $\frac{3}{4}$ of the previous layer in order to create the required bottleneck for the auto encoder. The size of the latent space was chosen to be 60. This choice was a good enough compromise between computational cost and reconstruction detail. Quadratic cost (MSE) was chosen as the cost function for its simplicity. The training was done for 5000 epochs with a batch size of 100 and a learning rate of 0.001. The sigmoid activation function was used for all layers. Standard MCMC methods were then used to generate the training data. For each temperature value, 10000 samples were generated and used as training data. The full code (with comments) for this paper can be found on github.com/BrandonWillnecker
\begin{acknowledgments}
	We would like to acknowledge the insightful comments and suggestions made by the referees. 
\end{acknowledgments}


\begin{thebibliography}{0}%
\makeatletter
\providecommand \@ifxundefined [1]{%
 \@ifx{#1\undefined}
}%
\providecommand \@ifnum [1]{%
 \ifnum #1\expandafter \@firstoftwo
 \else \expandafter \@secondoftwo
 \fi
}%
\providecommand \@ifx [1]{%
 \ifx #1\expandafter \@firstoftwo
 \else \expandafter \@secondoftwo
 \fi
}%
\providecommand \natexlab [1]{#1}%
\providecommand \enquote  [1]{``#1''}%
\providecommand \bibnamefont  [1]{#1}%
\providecommand \bibfnamefont [1]{#1}%
\providecommand \citenamefont [1]{#1}%
\providecommand \href@noop [0]{\@secondoftwo}%
\providecommand \href [0]{\begingroup \@sanitize@url \@href}%
\providecommand \@href[1]{\@@startlink{#1}\@@href}%
\providecommand \@@href[1]{\endgroup#1\@@endlink}%
\providecommand \@sanitize@url [0]{\catcode `\\12\catcode `\$12\catcode
  `\&12\catcode `\#12\catcode `\^12\catcode `\_12\catcode `\%12\relax}%
\providecommand \@@startlink[1]{}%
\providecommand \@@endlink[0]{}%
\providecommand \url  [0]{\begingroup\@sanitize@url \@url }%
\providecommand \@url [1]{\endgroup\@href {#1}{\urlprefix }}%
\providecommand \urlprefix  [0]{URL }%
\providecommand \Eprint [0]{\href }%
\providecommand \doibase [0]{https://doi.org/}%
\providecommand \selectlanguage [0]{\@gobble}%
\providecommand \bibinfo  [0]{\@secondoftwo}%
\providecommand \bibfield  [0]{\@secondoftwo}%
\providecommand \translation [1]{[#1]}%
\providecommand \BibitemOpen [0]{}%
\providecommand \bibitemStop [0]{}%
\providecommand \bibitemNoStop [0]{.\EOS\space}%
\providecommand \EOS [0]{\spacefactor3000\relax}%
\providecommand \BibitemShut  [1]{\csname bibitem#1\endcsname}%
\let\auto@bib@innerbib\@empty
\end{thebibliography}%


\newpage
\begin{thebibliography}{99}
	
\bibitem{CCC}
G. Carleo, I. Cirac, K. Cranmer, L. Daudet, M. Schuld, N. Tishby, L. Vogt-Maranto, and L. Zdeborová, Machine learning and the physical sciences, Rev. Mod. Phys. 91, 045002 (2019).

\bibitem{MLCM}
E. Bedolla, L. C. Padierna, and R. Castanneda-Priego, Machine learning for condensed matter physics, J. Phys.: Condens. Matter 33, 053001 (2021).

\bibitem{CM}
J. Carrasquilla and R. G. Melko, Machine learning phases of matter, Nat. Phys. 13, 431 (2017).

\bibitem{Wang}
L. Wang, Discovering phase transitions with unsupervised learning, Phys. Rev. B 94, 195105 (2016).

\bibitem{Loop}
Y. Zhang , E.-A Kim, Quantum Loop Topography for Machine Learning, Phys. Rev. Lett. 118, 216401 (2017)

\bibitem{Khan}
M. Richter-Laskowska, H. Khan, N. Trivedi, and M. M. Masa, A machine learning approach to the Berezinskii-Kosterlitz- Thouless transition in classical and quantum models, Condens. Matter Phys. 21, 33602 (2018)

\bibitem{Nieva}
J. F. Rodriguez-Nieva, M. S. Scheurer, Identifying topological order through unsupervised machine learning, Nat. Phys. 15, 790 (2019)

\bibitem{Beach}
M. J. S. Beach, A. Golubeva, and R. G. Melko, Machine learning vortices at the Kosterlitz-Thouless transition, Phys. Rev. B 97, 045207 (2018)

\bibitem{Rem}
B. S. Rem, N. Käming, M. Tarnowski, L. Asteria, N. Fläschner, C. Becker , K. Sengstock and C. Weitenberg, Identifying Quantum Phase Transitions using Artificial Neural Networks on Experimental Data, Nat. Phys. 15, 917 (2019)

\bibitem{Zhang}
W. Zhang, J. Liu, T. Wei, Machine learning of phase transitions in the percolation and XY models, Phys. Rev. E 99, 032142 (2019)

\bibitem{Ng}
K. Ng and M. Yang, Unsupervised learning of phase transitions via modified anomaly detection with autoencoders, Phys. Rev. B 108, 214428 (2023)

\bibitem{Shiina}
K. Shiina, H. Mori, Y. Okabe and H. K. Lee, Machine-Learning Studies on Spin Models, Sci Rep 10, 2177 (2020)

\bibitem{yusuke}
Y. Miyajima and M. Mochizuki, Machine-Learning Detection of the Berezinskii-Kosterlitz-Thouless Transition and the Second-Order Phase Transition in the XXZ models, Phys. Rev. B 107, 134420 (2023)

\bibitem{auto_endoder_info}
M. A. Kramer, Nonlinear principal component analysis using autoassociative neural networks, Aiche Journal 37, 233 (1991)

\bibitem{auto_encoder_use}
L. Theis, W. Shi, A. Cunningham and F. Huszár, Lossy Image Compression with Compressive Autoencoders, arXiv:1703.00395.

\bibitem{loss_fucntion}
K. Janocha and W. M. Czarnecki, On Loss Functions for Deep Neural Networks in Classification, arXiv:1702.05659

\bibitem{mnielson}
M. A. Nielsen, Neural Networks and Deep Learning, Determination Press (2015) 

\bibitem{Hohenberg_Mermin_Wagner_theorem}
N. D. Mermin and H. Wagner, Absence of Ferromagnetism or Antiferromagnetism in One- or Two-Dimensional Isotropic Heisenberg Models, Phys. Rev. Lett. 17, 1307 (1966)

\bibitem{xy1}
V. L. Berezinskii, Destruction of long range order in one- dimensional and two-dimensional systems having a continuous symmetry group. I. Classical systems, Sov. Phys. JETP 32, 493 (1971)

\bibitem{xy2}
V. L. Berezinsky, Destruction of long-range order in one- dimensional and two-dimensional systems possessing a contin- uous symmetry group. II. Quantum systems, Sov. Phys. JETP 34, 610 (1972)

\bibitem{xy3}
J. M Kosterlitz and D. J. Thouless, Ordering, metastability and phase transitions in two-dimensional systems, J. Phys. C 6, 1181 (1973)

\bibitem{xy4}
 J. M. Kosterlitz, The critical properties of the two-dimensional XY model, J. Phys. C 7, 1046 (1974)

\bibitem{vorext_density_sampling}
X. Leoncini, A. Verga and S. Ruffo, Hamiltonian Dynamics and the Phase Transition of the XY Model, Phys. Rev. E 57, 6377 (1998)


\bibitem{nn_symmetry}
B. Bloem-Reddy and Y. W. Teh, Probabilistic Symmetries and Invariant Neural Networks, Journal of Machine Learning Research 21, 1 (2020)

\bibitem{nn_symmetry_2}
J. Wood and J. Shawe-Taylor, Representation theory and invariant neural networks, Discrete Applied Maths 69, 33 (1996)
		
\bibitem{method_swendsen}
Swendsen, R.H. and Wang, J.S., 1987. Nonuniversal critical dynamics in Monte Carlo simulations. Physical review letters, 58(2), p.86.

\bibitem{method_wolff}
Wolff, U., 1989. Collective Monte Carlo updating for spin systems. Physical Review Letters, 62(4), p.361.

\bibitem{kt_temp}
Y. -D. Hsieh, Y. -J. Kao and A. W. Sandvik, Finite-size scaling method for the Berezinskii-Kosterlitz-Thouless transition, J. Stat. Mech., P09001 (2013)

\bibitem{intro_to_vae}
Diederik P. Kingma and Max Welling (2019), An Introduction to Variational Autoencoders, Now Foundations and Trends.

\bibitem{top_phase_trans}
M. H. Zarei (2019), Ising order parameter and topological phase transitions: Toric code in a uniform magnetic field, Phys. Rev. B 100, 125159 (2019)
\end{thebibliography}
\end{document}